\pdfoutput=1
\RequirePackage{ifpdf}
\ifpdf 
\documentclass[pdftex]{sigma}
\else
\documentclass{sigma}
\fi

\begin{document}

\allowdisplaybreaks

\renewcommand{\PaperNumber}{076}

\FirstPageHeading

\ShortArticleName{Ground-State Analysis for an Exactly Solvable Coupled-Spin Hamiltonian}

\ArticleName{Ground-State Analysis\\
for an Exactly Solvable Coupled-Spin Hamiltonian}

\Author{Eduardo MATTEI~$^\dag$ and Jon LINKS~$^\ddag$}

\AuthorNameForHeading{E.~Mattei and J.~Links}

\Address{$^\dag$~Centro Brasileiro de Pesquisas F\'{\i}sicas, Rua Dr.\ Xavier Sigaud 150, Rio de Janeiro, Brazil}
\EmailD{\href{mailto:eduardo.mattei@ufrgs.br}{eduardo.mattei@ufrgs.br}}

\Address{$^\ddag$~School of Mathematics and Physics, The University of Queensland, 4072, Australia}
\EmailD{\href{mailto:jrl@maths.uq.edu.au}{jrl@maths.uq.edu.au}}

\ArticleDates{Received July 23, 2013, in f\/inal form November 22, 2013; Published online November 30, 2013}

\Abstract{We introduce a~Hamiltonian for two interacting $\mathfrak{su}(2)$ spins.
We use a~mean-f\/ield analysis and exact Bethe ansatz results to investigate the ground-state properties of
the system in the classical limit, def\/ined as the limit of inf\/inite spin (or highest weight).
Complementary insights are provided through investigation of the energy gap, ground-state f\/idelity, and
ground-state entanglement, which are numerically computed for particular parameter values.
Despite the simplicity of the model, a~rich array of ground-state features are uncovered.
Finally, we discuss how this model may be seen as an analogue of the exactly solvable $p+ip$ pairing
Hamiltonian.
}


\Classification{81R05; 17B80; 81R12}

\section{Introduction}

We will analyse a  Hamiltonian for two interacting $\mathfrak{su}(2)$ spins, and begin by def\/ining the model.
The $\mathfrak{su}(2)$ algebra has generators $\{S^z, S^+, S^-\}$ with commutation relations
\begin{gather}
\left[S^+,S^-\right]=2S^z,
\qquad
\left[S^z,S^\pm\right]=\pm S^\pm.
\label{cr}
\end{gather}
Given a~f\/inite-dimensional module with highest weight $s$ ($2s\in {\mathbb N}\cup\{0\}$) the action of
the gene\-ra\-tors on the weight basis is
\begin{gather*}
S^z|s,m\rangle=m|s,m\rangle,
\\
S^+|s,m\rangle=\sqrt{(s-m)(s+m+1)}|s,m+1\rangle,
\\
S^-|s,m\rangle=\sqrt{(s+m)(s-m+1)}|s,m-1\rangle
\end{gather*}
with $m=-s,-s+1,\dots,s-1,s$.

We will study the following coupled-spin Hamiltonian
\begin{gather}
H=(1+G)\big(z_1^2\big(S_1^z+s\big)+z_2^2\big(S_2^z+s\big)\big)-G\big(z_1S_1^++z_2S_2^+\big)\big(z_1S_1^-+z_2S^-_2\big),
\label{H}
\end{gather}
which acts on the $(2s+1)^2$-dimensional tensor product space with basis
\begin{gather}
\{|s,m_1\rangle\otimes|s,m_2\rangle:\,m_j=-s,-s+1,\dots,s-1,s;\,j=1,2\}.
\label{basis}
\end{gather}
The Hamiltonian commutes with $S^z=S^z_1+S^z_2$.
Throughout we assume $z_1> z_2\geq 0$.
For given~$s$ the model has three coupling parameters.
One of these can be f\/ixed by choosing the overall energy scale, e.g.,
setting $z_1=1$, which leaves two independent parameters $z_2$ and $G$.
Our main objective is to investigate ground-state properties of the model in the {\it classical limit}
which we def\/ine by a~sequence of ordered triples $(G,s,m)$ such that $G\rightarrow 0$, $s
\rightarrow\infty$, $m\rightarrow\pm\infty$ (or $m=0$) with $\mu=m/(2s)$ and $g=4sG$.

An advantage of this model is that analytic results are accessible.
Despite the simplicity of the Hamiltonian, a~rich array of ground-state features are uncovered.
In Section~\ref{mfa} we conduct an analysis of the system using a~mean-f\/ield approximation.
In contrast, Section~\ref{exactsol} is devoted to the exact calculation of certain properties,
including results obtained from an exact Bethe ansatz solution of the Hamiltonian, in order to obtain the
ground-state phase diagram.
In Section~\ref{numeric} we numerically analyse the system to study the ground-state energy gap,
f\/idelity~\cite{zp06,zb07}, and entanglement~\cite{on02,oaff02}, which have been used as indicators of
quantum phase transitions in other simple bipartite quantum systems, e.g.,~\cite{hmm03,pd05,rfmr12}.
Conclusions are drawn in Section~\ref{conc}.

The model considered here may be viewed as a~spin-analogue of the exactly solvable $p+ip$ pairing
Hamiltonian~\cite{dilsz10,ilsz09,rdo10}.
The Hamiltonian came to some prominence through the work of Volovik~\cite{v} and Read and Green~\cite{rg00}
who exposed a~change in the topological properties of the ground-state wavefunction of the Hamiltonian as
a~function of the coupling parameter.
This result has generated many subsequent studies, as the topological nature of the transition is quite
distinctive from conventional forms of phase transition characterised by an order parameter.
In the Conclusion we will provide further discussion on this point and the relation between the
coupled-spin model and the $p+ip$ model.

We also remark that a~related study was recently given in~\cite{ld13} in connection with the
Lipkin--Meshkov--Glick model.
A signif\/icant dif\/ference however is that~\cite{ld13} realises the Hamiltonian with
inf\/inite-dimensional unitary representations of coupled $\mathfrak{su}(1,1)$ spins, in contrast to our approach
using f\/inite-dimensional unitary representations of the coupled $\mathfrak{su}(2)$ spins.

\section{Mean-f\/ield analysis}\label{mfa}

The Hamiltonian~\eqref{H} can be approximated by a~Hamiltonian with linear terms through
\begin{gather*}
AB\approx A\langle B\rangle+\langle A\rangle B-\langle A\rangle\langle B\rangle,
\end{gather*}
where $\langle A \rangle $ denotes the expectation value of the operator~$A$.
The mean-f\/ield approximation which linearises the Hamiltonian~\eqref{H} is
\begin{gather*}
H_{\rm mf}=(1+G)s\big(z_1^2+z_2^2\big)+\rho m+\big((1+G)z_1^2-\rho\big)S_1^z+\big((1+G)z_2^2-\rho\big)S_2^z
\\
\phantom{H_{\rm mf}=}{}
-\frac{\hat{\Delta}}{{2}}\big(z_1S_1^++z_2S_2^+\big)-\frac{\hat{\Delta}^*}{{2}}\big(z_1S_1^-+z_2S_2^-\big)+\frac{\Delta^2}{4G},
\end{gather*}
where $\rho$ is a~Lagrange multiplier, which is introduced since the mean-f\/ield approximation breaks the
conservation of $S^z$, and
\begin{gather*}
\hat{\Delta}=2{G}\langle z_1S_1^-+z_2S_2^-\rangle.
\end{gather*}
Setting $\Delta = |\hat{\Delta} |$ the ground state energy is found to be
\begin{gather*}
{E}_0=(1+G)s\big(z_1^2+z_2^2\big)+\rho m+\frac{\Delta^2}{4G}-s\sqrt{\big((1+G)z_1^2-\rho\big)^2+z_1^2\Delta^2}
\\
\phantom{{E}_0=}{}
-s\sqrt{\big((1+G)z_2^2-\rho\big)^2+z_2^2\Delta^2}.
\end{gather*}
Using the Hellmann--Feynman theorem, we can determine mean-f\/ield expectation values:
\begin{gather*}
\left<\frac{\partial H_{\rm mf}}{\partial G}\right>=\frac{\partial{E}_0}{\partial G},
\end{gather*}
which leads to
\begin{gather*}
z_1^2\langle S_1^z\rangle+z_2^2\langle S_2^z\rangle+\frac{\Delta^2}{2G^2}
=sz_1^2\frac{(1+G)z_1^2\!-\rho+\Delta^2G^{-1}}{\sqrt{((1+G)z_1^2\!-\rho)^2+z_1^2\Delta^2}}
+sz_2^2\frac{(1+G)z_2^2\!-\rho+\Delta^2G^{-1}}{\sqrt{((1+G)z_2^2\!-\rho)^2+z_2^2\Delta^2}}.
\end{gather*}
Taking the classical limit gives
\begin{gather}
\frac{2}{g}=\frac{z_1^2}{\sqrt{(z_1^2-\rho)^2+z_1^2\Delta^2}}+\frac{z_2^2}
{\sqrt{(z_2^2-\rho)^2+z_2^2\Delta^2}}.
\label{gap}
\end{gather}
Also
\begin{gather*}
\left<\frac{\partial H_{\rm mf}}{\partial\rho}\right>=\frac{\partial E_0}{\partial\rho}
\end{gather*}
yielding
\begin{gather*}
-\langle S_1^z\rangle-\langle S_2^z\rangle=\frac{s((1+G)z_1^2-\rho)}
{\sqrt{((1+G)z_1^2-\rho)^2+z_1^2\Delta^2}}+\frac{s((1+G)z_2^2-\rho)}
{\sqrt{((1+G)z_2^2-\rho)^2+z_2^2\Delta^2}}.
\end{gather*}
Taking the classical limit and using~\eqref{gap} gives
\begin{gather}
\frac{2}{g}+2\mu=\frac{\rho}{\sqrt{(z_1^2-\rho)^2+z_1^2\Delta^2}}+\frac{\rho}
{\sqrt{(z_2^2-\rho)^2+z_2^2\Delta^2}}.
\label{chempot}
\end{gather}
Returning to the energy expression in the limit $s{\rightarrow}\infty$, and taking into account
equa\-tions~\eqref{gap},~\eqref{chempot}, we obtain the intensive energy expression
\begin{gather}
e_0=\lim_{s\rightarrow\infty}\frac{{E}_0}{s}
=\big(z_1^2+z_2^2\big)+2\rho\mu+\frac{\Delta^2}{g}-\sqrt{(z_1^2-\rho)^2+z_1^2\Delta^2}
-\sqrt{(z_2^2-\rho)^2+z_2^2\Delta^2}
\nonumber
\\
\phantom{e_0}{}
=z_1^2\left(1-\frac{1}{2}\frac{2z_1^2+\Delta^2-2\rho}{\sqrt{(z_1^2-\rho)^2+z_1^2\Delta^2}}
\right)+z_2^2\left(1-\frac{1}{2}\frac{2z_2^2+\Delta^2-2\rho}{\sqrt{(z_2^2-\rho)^2+z_2^2\Delta^2}}\right).
\label{mfnrg}
\end{gather}
It may be verif\/ied that setting $\rho=\Delta^2/4$ in~\eqref{gap}, \eqref{chempot} and combining these
equations leads to
\begin{gather}
\mu=1-2g^{-1}.
\label{mrl}
\end{gather}
Furthermore, setting $\rho=\Delta^2/4$ in~\eqref{mfnrg} gives $e_0=0$.

Recalling the assumption $z_1 > z_2$ the minimum excitation energy (i.e.\
the gap) is
\begin{gather}
\Delta E=\sqrt{\big(z_2^2-\rho\big)^2+z_2^2\Delta^2},
\label{nrggap}
\end{gather}
which vanishes when $z_2=\rho=0$.
However turning to~\eqref{chempot} we see that the limits $z_2\rightarrow 0$ and $\rho\rightarrow 0$ do not
commute in that equation.
First setting $\rho=0$ leads to
\begin{gather}
\mu=-g^{-1}.
\label{rgl}
\end{gather}
Alternatively f\/irst setting $z_2=0$ yields
\begin{gather}
\mu=\frac{1}{2}\lim_{\rho\rightarrow0}\frac{\rho}{|\rho|}-\frac{1}{g},
\label{mub}
\end{gather}
which shows a~discontinuity as $\rho$ passes through zero.
This indicates that while gapless excitations occur in the limit $z_2,\rho\rightarrow 0$, where they are to
be found in the phase diagram will depend on how these two limits are taken (at least within the
mean-f\/ield approximation).
A~unusual property is that equations~\eqref{mrl},~\eqref{rgl},~\eqref{mub} do not depend on~$z_1$ and~$z_2$.
To gain a~better appreciation of these features we next conduct an analysis using exact results.

\section{The exact solution}
\label{exactsol}
We start with the observation that
\begin{gather*}
H\left(|s,-s\rangle\otimes|s,-s\rangle\right)=0.
\end{gather*}
To determine exact eigenstates of~\eqref{H} by way of a~Bethe ansatz solution, we def\/ine generic states
of the form
\begin{gather*}
|\psi\rangle=\prod_{k=1}^M C(y_k)\left(|s,-s\rangle\otimes|s,-s\rangle\right),
\\
|\psi_j\rangle=\prod_{k\neq j}^M C(y_k)\left(|s,-s\rangle\otimes|s,-s\rangle\right),
\\
|\psi_{jl}\rangle=\prod_{k\neq j,l}^M C(y_k)\left(|s,-s\rangle\otimes|s,-s\rangle\right),
\end{gather*}
where
\begin{gather*}
C(y)=\frac{z_1S_1^+}{y-z_1^2}+\frac{z_2S_2^+}{y-z_2^2}.
\end{gather*}
We also def\/ine
\begin{gather*}
H_0=z_1^2\big(S_1^z+s\big)+z_2^2\big(S_2^z+s\big),
\qquad
Q^+=z_1S_1^++z_2S_2^+,
\qquad
Q^-=z_1S^-_1+z_2S^-_2,
\end{gather*}
such that
\begin{gather*}
H=(1+G)H_0-G Q^+Q^-,
\end{gather*}
and note the following commutation relations:
\begin{gather*}
\left[H_0, C(y)\right]=\frac{z^3_1}{y-z_1^2}S_1^++\frac{z^3_2}{y-z_2^2}S_2^+=yC(y)-Q^+,
\\
\left[Q^-, C(y)\right]=-2\frac{z_1^2}{y-z_1^2}S_1^z-2\frac{z_2^2}{y-z_2^2}S_2^z.
\end{gather*}
By direct calculation it is then found that
\begin{gather*}
H|\Psi\rangle=(1+G)H_0|\Psi\rangle-G Q^+Q^-|\Psi\rangle
=(1+G)\sum_{j=1}^M\left(y_j|\Psi\rangle-Q^+|\Psi_j\rangle\right)
\\
\phantom{H|\Psi\rangle=}
{}+2G Q^+\sum_{j=1}^M\sum_{p=1}^2\left(C(y_1)\cdots\left(\frac{z_p^2}{y-z_p^2}S_p^z\right)\cdots C(y_M)\right)|0\rangle
\\
\phantom{H|\Psi\rangle}{}
=(1+G)\sum_{j=1}^M\left(y_j|\Psi\rangle-Q^+|\Psi_j\rangle\right)+2G Q^+\sum_{j=1}^M\left(\frac{z_1^2s}
{y_j-z_1^2}+\frac{z_2^2s}{y_j-z_2^2}\right)|\Psi_j\rangle
\\
\phantom{H|\Psi\rangle=}
{}+2G Q^+\sum_{j=1}^M\sum_{r>j}^M\sum_{p=1}^2\frac{z_p^3}{(y_j-z_p^2)(y_r-z_p^2)}S_p^+|\Psi_{rj}\rangle
\\
\phantom{H|\Psi\rangle}
=(1+G)\sum_{j=1}^M\left(y_j|\Psi\rangle-Q^+|\Psi_j\rangle\right)-2G Q^+\sum_{j=1}^M\left(\frac{z_1^2s}
{y_j-z_1^2}+\frac{z_2^2s}{y_j-z_2^2}\right)|\Psi_j\rangle
\\
\phantom{H|\Psi\rangle=}
{}+G Q^+\sum_{j=1}^M\sum^M_{r\neq j}\sum_{p=1}^2\left(\frac{z_p y_r}{(y_j-y_r)(y_r-z_p^2)}+\frac{z_py_j}
{(y_r-y_j)(y_j-z_p^2)}\right)S_p^+|\Psi_{rj}\rangle
\\
\phantom{H|\Psi\rangle}{}
=(1+G)\sum_{j=1}^M\left(y_j|\Psi\rangle-Q^+|\Psi_j\rangle\right)-2G Q^+\sum_{j=1}^M\left(\frac{z_1^2s}
{y_j-z_1^2}+\frac{z_2^2s}{y_j-z_2^2}\right)|\Psi_j\rangle
\\
\phantom{H|\Psi\rangle=}
{}+G Q^+\sum_{j=1}^M\sum^M_{r\neq j}\left(\frac{y_r}{y_j-y_r}|\Psi_j\rangle+\frac{y_j}{y_r-y_j}
|\Psi_r\rangle\right)
\\
\phantom{H|\Psi\rangle}{}
=(1+G)\sum_{j=1}^M\left(y_j|\Psi\rangle-Q^+|\Psi_j\rangle\right)-2G Q^+\sum_{j=1}^M\left(\frac{z_1^2s}
{y_j-z_1^2}+\frac{z_2^2s}{y_j-z_2^2}\right)|\Psi_j\rangle
\\
\phantom{H|\Psi\rangle=}
{}+2G Q^+\sum_{j=1}^M\sum_{r\neq j}^M\frac{y_r}{y_j-y_r}|\Psi_j\rangle
\\
\phantom{H|\Psi\rangle}{}
=(1+G)\sum_{j=1}^M\left(y_j|\Psi\rangle-Q^+|\Psi_j\rangle\right)-2G Q^+\sum_{j=1}^M\left(\frac{z_1^2s}
{y_j-z_1^2}+\frac{z_2^2s}{y_j-z_2^2}\right)|\Psi_j\rangle
\\
\phantom{H|\Psi\rangle=}
{}+2G Q^+\sum_{j=1}^M\sum^M_{r\neq j}\frac{y_r}{y_j-y_r}|\Psi_j\rangle.
\end{gather*}
The terms proportional to $|\Psi_j\rangle$ cancel provided
\begin{gather*}
(1+G)+2G\left(\frac{z_1^2s}{y_j-z_1^2}+\frac{z_2^2s}{y_j-z_2^2}\right)=2G\sum^M_{r\neq j}\frac{y_r}
{y_j-y_r},
\qquad
k=1,\dots,M,
\end{gather*}
which can be equivalently written as
\begin{gather}
\frac{G^{-1}+2M+4s-1}{y_k}+\frac{2s}{y_k-z_1^2}+\frac{2s}{y_k-z_2^2}=\sum^M_{j\neq k}\frac{2}{y_k-y_j},
\qquad
k=1,\dots,M.
\label{bae}
\end{gather}
For each solution of the coupled equations~\eqref{bae}, $|\Psi\rangle$ is an eigenstate of~\eqref{H} with
energy eigenvalue given by
\begin{gather*}
E=(1+G)\sum_{k=1}^M y_k.
\end{gather*}

\subsection{Duality relations}
\label{dual}

For $z_2\neq 0$, from the commutation relations~\eqref{cr} the following equation can be shown to hold
using proof by induction:
\begin{gather}
\big[H, (C(0))^M\big]=-M Q^+(C(0))^{M-1}\big(1+GM+2GS^z\big).
\label{commu}
\end{gather}
If a~state $ |\psi\rangle $ has spin $m$ (the eigenvalue of $S^z$) the above equation shows that choosing
\begin{gather*}
1+GM+2Gm=0,
\end{gather*}
so $M=-2m-G^{-1}$, the two states $ |\psi\rangle $ and $|\tilde{\psi}\rangle = (C(0))^{M}|\psi\rangle$ have
the same energy.
We call the states $|\psi\rangle $ and $|\tilde{\psi}\rangle$ {\it dual} states.
Since the state $ |\psi\rangle $ has spin $m$, the state $|\tilde{\psi}\rangle $ has spin $m+M=-m-G^{-1}$.
We identify the following special cases:
\begin{itemize}\itemsep=0pt
\item The dual of the lowest weight state $|\psi\rangle=|s,-s\rangle \otimes |s,-s\rangle$ occurs when
$M=4s-G^{-1}$ leading to $|\tilde{\psi}\rangle $ having spin $m= 2s-G^{-1}$.
In the classical limit $s\rightarrow\infty$ this yields equation~\eqref{mrl}, for which the ground-state
energy is zero.
Explicitly
\begin{gather}
|\tilde{\psi}\rangle=(C(0))^M|s,-s\rangle\otimes|s,-s\rangle.
\label{mrstate}
\end{gather}
\item If $M=0$ then a~state is self-dual when $m=-(2G)^{-1}$.
In the classical limit $s\rightarrow\infty$ this yields equation~\eqref{rgl}.
\end{itemize}

\begin{figure}[t] \centering
\includegraphics[height=10cm,angle=-90]{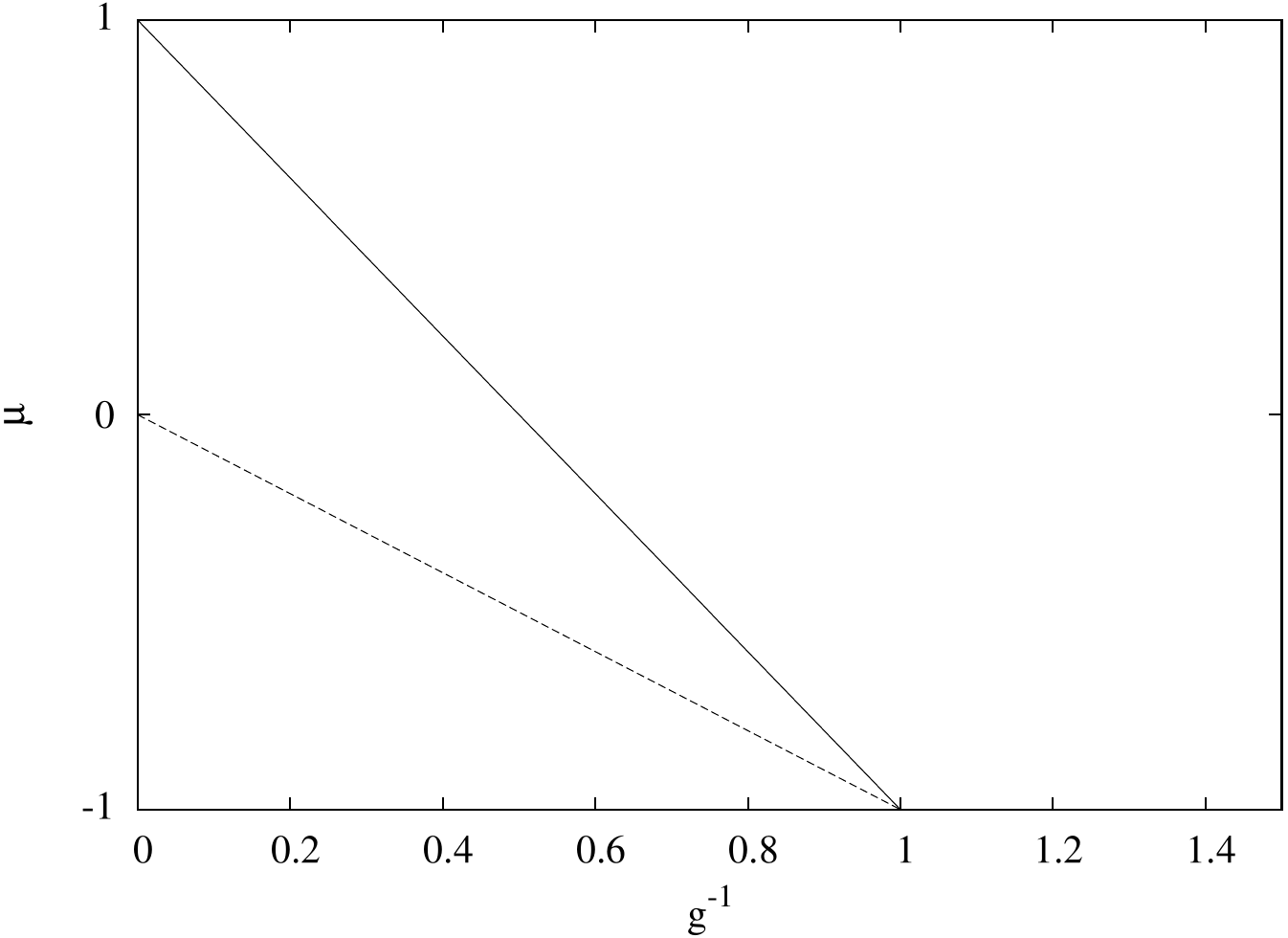}

\caption{Duality diagram.
The solid line represents equation~\eqref{mrl}, on which the intensive ground-state energy is zero.
The dotted line represents equation~\eqref{rgl}.
The mean-f\/ield analysis predicts that there are gapless excitations {\it on} equation~\eqref{rgl} as
$z_2\rightarrow 0$, but also that the limits $\mu\rightarrow -g^{-1}$ and $z_2\rightarrow 0$ do not commute
in equation~\eqref{chempot}.}
\label{fig1}
\end{figure}

Note that neither line given by~\eqref{mrl} nor~\eqref{rgl} depends on the variables $z_1$ and $z_2$.
However, as already seen in the mean-f\/ield analysis, caution must be taken when considering the limit
$z_2\rightarrow 0$.
Since $C(0)$ is singular in the limit $z_2\rightarrow 0$, we def\/ine rescaled operators
\begin{gather*}
\tilde{C}(0)=z_1z_2C(0)=z_2S_1^++z_1S^+_2.
\end{gather*}
Equation~\eqref{commu} assumes the form
\begin{gather}
\big[H, (\tilde{C}(0))^M\big]=-z_1z_2M Q^+\big(\tilde{C}(0)\big)^{M-1}\big(1+GM+2GS^z\big).
\label{magic}
\end{gather}
It is easily recognized that the commutator vanishes in the limit $z_2\rightarrow 0$.
In this case the classes of states with the same energy generally have greater cardinality than those of
the dual state classes identif\/ied above for $z_2\neq 0$.

\subsection{Classical limit of the ground-state exact solution}
Following \cite{dilsz10,ml12,rdo10}, we look to
solve the Bethe ansatz equations in the classical limit under the assumption that in this limit
the roots of~\eqref{bae} are densely distributed on a curve $\Gamma$ in the complex plane
which is invariant under ref\/lection about the real axis.
Introducing $r(y)$ as the density function on $\Gamma$ for the roots, we divide~\eqref{bae} by~$2s$ and
take the limit $s \to \infty$ to obtain
\begin{gather}
\frac{2g^{-1}+2\mu}{y}+\frac{1}{y-z^2_1}+\frac{1}{y-z^2_2}=2P\int_{\Gamma}|dy'|\frac{r(y')}{y-y'},
\label{limit}
\end{gather}
where $P$ refers to the Cauchy principal value for the integral.
To solve the problem we def\/ine an analytic function $h(y')$ outside $\Gamma$ such that
\begin{gather*}
r(y')=h_+(y')-h_-(y'),
\end{gather*}
where $h_+(y')$ and $h_-(y')$ are the limiting values of $h(y')$ to the right and left of $\Gamma$.
We can then rewrite~\eqref{limit} as
\begin{gather}
\frac{2g^{-1}+2\mu}{y}+\frac{1}{y-z^2_1}+\frac{1}{y-z^2_2}=2\oint_{L}dy'\frac{h(y')}{y-y'},
\label{lim2}
\end{gather}
where $L$ encircles $\Gamma$.
Next we take the following ansatz:
\begin{gather}
h(y)=\frac{R(y)}{2\pi i}\left[\frac{\phi(z_1)}{z_1^2-y}+\frac{\phi(z_2)}{z^2_2-y}+\frac{S}{y}\right],
\label{hy}
\end{gather}
where
\begin{gather*}
R(y)=\sqrt{(y-a)(y-b)}
\end{gather*}
with $a$, $b=a^*$ denoting the end points of the ${\rm arc}\,\Gamma$.
Substituting~\eqref{hy} into~\eqref{lim2} and evaluating the contour integral by residues leads to the
conditions
\begin{gather*}
\phi(z_j)=\frac{1}{2\sqrt{(z_j^2-a)(z_j^2-b)}},
\qquad
S=\phi(z_1)+\phi(z_2),
\qquad
S=\frac{g^{-1}+\mu}{\sqrt{ab}}.
\end{gather*}
These combine to give
\begin{gather}
\frac{2}{g}+2\mu=\frac{\sqrt{ab}}{\sqrt{(z^2_1-a)(z^2_1-b)}}+\frac{\sqrt{ab}}{\sqrt{(z^2_2-a)(z^2_2-b)}}.
\label{q}
\end{gather}
We may also compute
\begin{gather}
\mu=-1+\int_{\Gamma}|dy'|\,r(y')=-1+\oint_L dy'\,h(y')
\nonumber
\\
\phantom{\mu}=\left(\frac{a+b}{2}\right)S-z_1^2\phi(z_1)-z_2^2\phi(z_2)-\frac{(\sqrt{a}-\sqrt{b})^2}{2}S
\nonumber
\\
\Longrightarrow
\quad
\frac{2}{g}=\frac{z^2_1}{\sqrt{(z^2_1-a)(z^2_1-b)}}+\frac{z^2_2}{\sqrt{(z^2_2-a)(z^2_2-b)}}.
\label{M}
\end{gather}

Identifying $\rho=\sqrt{ab}$ and $\Delta^2=a+b+2\sqrt{ab}$, equations~\eqref{q},~\eqref{M} are precisely
the mean-f\/ield equations~\eqref{chempot},~\eqref{gap} respectively.
This would suggest that the mean-f\/ield equations should be exact in the classical limit as they are
reproducible from the exact Bethe ansatz solution.
However, the above calculations assume that in the classical limit the roots of the Bethe ansatz equations
become dense on a~curve $\Gamma$.
This assumption is not always valid.
Suppose that the curve $\Gamma$ closes such that the end points $a$, $b$ become real and equal.
Then equations~\eqref{q},~\eqref{M} reduce to
\begin{gather*}
\frac{2}{g}+2\mu=\frac{|a|}{|z^2_1-a|}+\frac{|a|}{|z^2_2-a|},
\qquad
\frac{2}{g}=\frac{z^2_1}{|z^2_1-a|}+\frac{z^2_2}{|z^2_2-a|}.
\end{gather*}
If $a\geq 0$ the only solution is $\mu=-1$, whereas if $a<0$ we obtain equation~\eqref{mrl}.
But the ground-state roots in this instance are all zero, as the ground state is dual to the lowest weight
state as shown in Subsection~\ref{dual} (see equation~\eqref{mrstate}).
Thus taking $s\rightarrow \infty$ after $g\rightarrow 2(1-\mu)^{-1}$ leads to a~delta function distribution
of roots at the origin, while taking $g\rightarrow 2(1-\mu)^{-1}$ from below after $s\rightarrow \infty$
leads to the roots being distributed on a~closed curve enclosing the origin.
We may identify equation~\eqref{mrl} as the boundary at which mean-f\/ield results break down.
In the region immediately beyond the line given by equation~\eqref{mrl}, we cannot assume that the
ground-state roots of the Bethe ansatz equations are densely distributed on a~single curve in order to
produce the mean-f\/ield equations from the Bethe ansatz solution.
A numerical study of this case has been undertaken in~\cite{ml12}.

\subsection[The limiting case $z_2=0$]{The limiting case $\boldsymbol{z_2=0}$}

In view of equation~\eqref{nrggap}, which indicates that the minimum excitation energy approaches zero as
$z_2\rightarrow 0$, here we set $z_2=0$ from the outset.
We also f\/ix the energy scale by letting $z_1=1$, in which case the Hamiltonian is simply
\begin{gather}
H=(1+G)\big(S_1^z+s\big)-G S_1^+S_1^-.
\label{degenham}
\end{gather}
The above Hamiltonian is diagonal in the basis~\eqref{basis} with eigenvalues
\begin{gather*}
E=m_1+s-G\big(s^2-m_1^2\big).
\end{gather*}

For each value of $m=m_1+m_2$ we can determine the ground state energy, and the gap to the f\/irst excited
state, by considering where level crossings occur.
We need to consider several dif\/ferent cases.
Starting with $m=m_1+m_2<0$ we have
\begin{gather*}
m_1=-s,
\quad
m_2=m+s
\quad
\Longrightarrow
\quad
E=0,
\\
m_1=-s+1,
\quad
m_2=m+s-1
\quad
\Longrightarrow
\quad
E=(1+G)-2sG.
\end{gather*}
In the classical limit it is seen that that for $m<0$ a~f\/irst level crossing occurs at
\begin{gather}
g=2.
\label{b1}
\end{gather}
Next consider
\begin{gather*}
m_1=s+m-1,
\quad
m_2=-s+1
\quad
\Longrightarrow
\quad
E=2s+m-1+G\big((m-1)^2+2s(m-1)\big),
\\
m_1=s+m,
\quad
m_2=-s
\quad
\Longrightarrow
\quad
E=2s+m+G\big(m^2+2sm\big).
\end{gather*}
Here there is an energy gap
\begin{gather*}
\Delta E=|1+G(2m-1+2s)|.
\end{gather*}
In the classical limit the f\/inal level crossing occurs when
\begin{gather}
\mu=-\frac{1}{2}-\frac{1}{g},
\label{b2}
\end{gather}
which requires that $\mu<-1/2$.

\begin{figure}[t] \centering
\includegraphics[height=10cm,angle=-90]{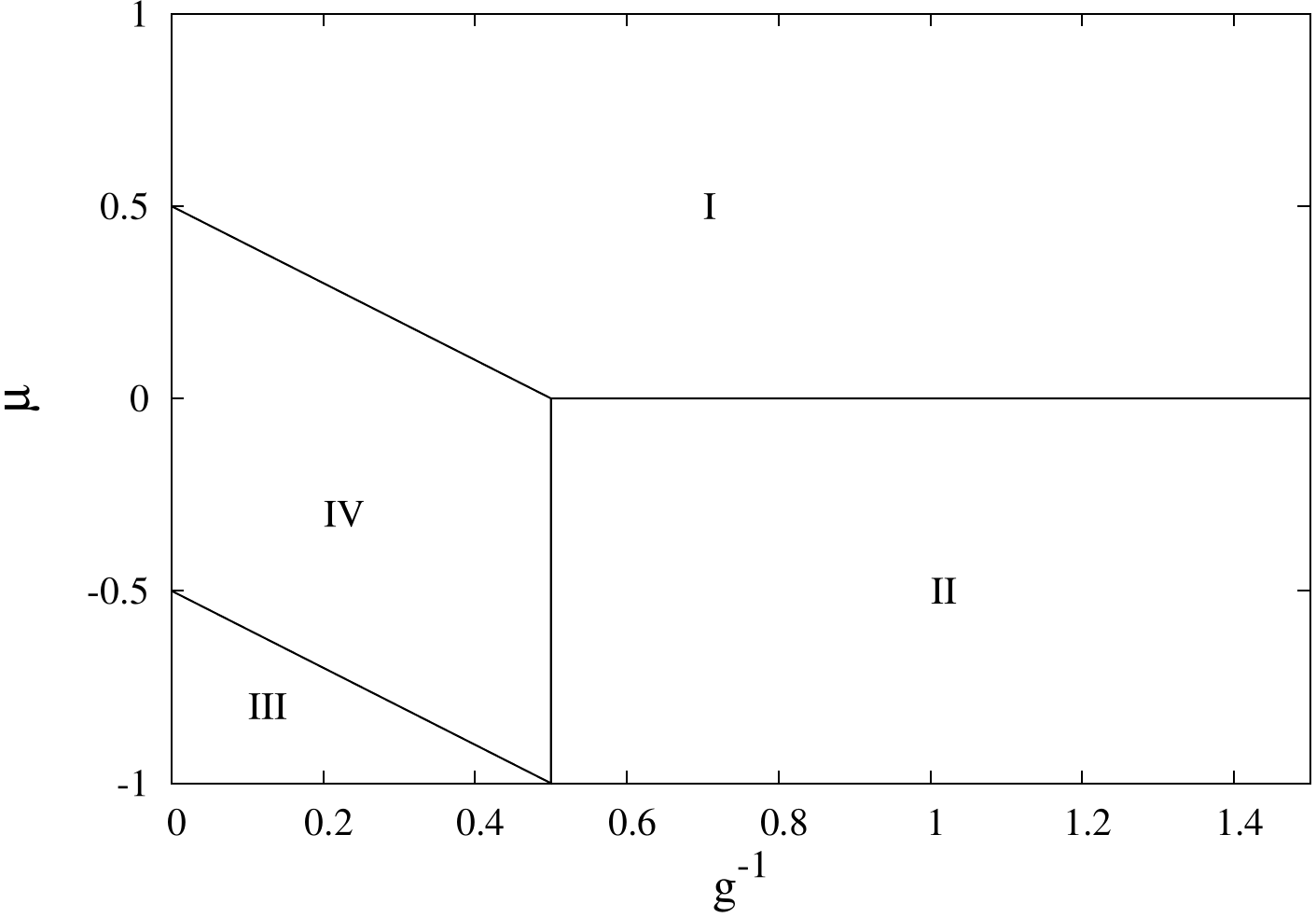}

\caption{Order parameter phase diagram.
Expressions for the gap between the ground state and f\/irst excited state for $z_2=0$ in the classical
limit are given by equations~\eqref{gap1}--\eqref{gap4}, and expressions for the order parameter~\eqref{op}
are given by equations~\eqref{op1}--\eqref{op4}.
Region IV is gapless, and is the only region in which the order parameter varies as a~function of $g$.
Note that the boundary lines between regions I and IV, and also~III and IV, are given by~\eqref{mub} for
$\rho>0$ and $\rho<0$ respectively.
Neither lines given by equation~\eqref{mrl} nor equation~\eqref{rgl} of Fig.~\ref{fig1} are present in the
diagram.}
\label{phasez}
\label{fig2}
\end{figure}

Now we consider the case for $m\geq 0$:
\begin{gather*}
m_1=-s+m,\quad m_2=s\quad\Longrightarrow\quad E=m+G\big(m^2-2sm\big),
\\
m_1=-s+m+1,\quad m_2=s-1\quad\Longrightarrow\quad E=m+1+G\big((m+1)^2-2s(m+1)\big).
\end{gather*}
The gap
\begin{gather*}
\Delta E=|1+G(2m+1-2s)|
\end{gather*}
goes to zero in the classical limit when
\begin{gather}
\mu=\frac{1}{2}-\frac{1}{g}.
\label{b3}
\end{gather}
This requires that $0\leq\mu<1/2$.
In the region bounded by the equations~\eqref{b1}, \eqref{b2}, \eqref{b3} the level crossings become dense in
the classical limit, giving rise to a~gapless phase.
The following relations hold for the gap between the ground state and f\/irst excited state in the regions
depicted in the phase diagram of Fig.~\ref{fig2}:
\begin{alignat}{3}
& {\rm I}:
\quad &&
\Delta E=1+g(\mu-1/2),&
\label{gap1}
\\
& {\rm II}:
\quad &&
\Delta E=1-g/2,&
\label{gap2}
\\
& {\rm III}:
\quad &&
\Delta E=-1-g(\mu+1/2),&
\label{gap3}
\\
& {\rm IV}:
\quad &&
\Delta E=0.&
\label{gap4}
\end{alignat}

Moreover, when $z_2=0$ the Hamiltonian~\eqref{degenham} acquires an $\mathfrak{su}(2)$ symmetry due to commutativity
with the generators $\{S_2^z, S_2^+, S_2^-\}$.
In view of this, we def\/ine an order parameter
\begin{gather}
{\mathcal O}=\lim_{s\rightarrow\infty}\frac{\left<S_2^z\right>}{s}.
\label{op}
\end{gather}
Analogous to the procedure described above, the following relations are found for the order parameter in
the regions depicted in the phase diagram of Fig.~\ref{fig2}:
\begin{alignat}{3}
& {\rm I}:
\quad &&
{\mathcal O}=1, &
\label{op1}
\\
& {\rm II}:
\quad &&
{\mathcal O}=1+2\mu, &
\label{op2}
\\
& {\rm III}:
\quad &&
{\mathcal O}=-1,&
\label{op3}
\\
& {\rm IV}:
\quad &&
{\mathcal O}=2\big(\mu+g^{-1}\big).&
\label{op4}
\end{alignat}
Only in the gapless region IV does the order parameter vary as a~function of~$g$.
The boundary lines between the regions are associated with discontinuities in the partial derivatives of
${\mathcal O}$.

There is clear discrepancy between Figs.~\ref{fig1} and~\ref{fig2} with respect to the vanishing of the
energy gap.
In deriving the two diagrams we consider behaviours as $G\rightarrow 0$, $s\rightarrow \infty$,
$m\rightarrow \infty$ with $g=4sG$, $\mu=m(2s)^{-1}$, and as $z_2\rightarrow 0$, yielding phase boundary
lines as functions of the coupling parameter~$g$.
In the next subsection we provide calculations which support the phase diagram shown in Fig.~\ref{fig2}.

\subsection[The limiting case $\mu= -1$]{The limiting case $\boldsymbol{\mu= -1}$}
\label{muminusone}

In the sector where $\mu=-1+(2s)^{-1}$, or equivalently $m=1-2s$, the Hilbert space is two-dimensional and
exact results are easily extracted.
In matrix form, with respect to the basis
\begin{gather*}
\{|s,-s+1\rangle\otimes|s,-s\rangle,\,|s,-s\rangle\otimes|s,-s+1\rangle\},
\end{gather*}
the Hamiltonian is
\begin{gather*}
H=\left(
\begin{matrix}
 (1+G-2sG)z_1^2&-2sGz_1z_2
\\
-2sGz_1z_2&(1+G-2sG)z_2^2
\end{matrix}
\right)
\end{gather*}
with the ground-state energy
\begin{gather*}
E=\frac{(1-g/2+g(4s)^{-1})(z_1^2+z_2^2)}{2}-\frac{1}{2}\sqrt{(1-g/2+g(4s)^{-1}
)^2\big(z_1^2-z_2^2\big)^2+g^2z_1^2z_2^2}
\end{gather*}
and energy gap to the excited state
\begin{gather}
\Delta E=\sqrt{\big(1-g/2+g(4s)^{-1}\big)^2\big(z_1^2-z_2^2\big)^2+g^2z_1^2z_2^2}.
\label{egap}
\end{gather}

The mean-f\/ield results of Section~\ref{mfa} indicate that the vanishing of the energy gap depends
on how the limits $z_2\rightarrow 0$ and $\rho \rightarrow 0$ are taken.
It is found from~\eqref{egap} that
\begin{gather*}
\lim_{g\rightarrow2}\lim_{z_2\rightarrow0}\lim_{s\rightarrow\infty}\Delta E=0
\end{gather*}
with the result being invariant with respect to interchanging the order of the limits.
Moreover, as $s\rightarrow\infty$ the {\it only} solution for $\Delta E =0$ when $z_1\neq 0$ is given by
$z_2=0$, $g=2$, meaning that gapless excitations {\it do not} occur on the line given by
equation~\eqref{rgl} in the limit $\mu\rightarrow -1$.

Next consider
\begin{gather*}
\frac{\partial(\Delta E)}{\partial g}
=\frac{g\big(z_1^2+z_2^2\big)^2+\big({-}2+(1-g)s^{-1}+g(4s^2)^{-1}\big)
\big(z_1^2-z_2^2\big)^2}{4\Delta E}.
\end{gather*}
We then have
\begin{gather*}
\lim_{s\rightarrow\infty}\lim_{z_2\rightarrow0}\lim_{g\rightarrow2}
\frac{\partial(\Delta E)}{\partial g}\\
\qquad{}
=\lim_{s\rightarrow\infty}\lim_{z_2\rightarrow0}\lim_{g\rightarrow2}
\frac{g\big(z_1^2+z_2^2\big)^2+\big({-}2+(1-g)s^{-1}+g(4s^2)^{-1}\big)
\big(z_1^2-z_2^2\big)^2}{4\sqrt{(1-g/2+g(4s)^{-1})^2(z_1^2-z_2^2)^2+g^2z_1^2z_2^2}}
\\
\qquad{}
=\lim_{s\rightarrow\infty}\lim_{z_2\rightarrow0}
\frac{2\big(z_1^2+z_2^2\big)^2+\big({-}2-s^{-1}+(2s^2)^{-1}\big)
\big(z_1^2-z_2^2\big)^2}{4\sqrt{((2s)^{-2}(z_1^2-z_2^2)^2+4z_1^2z_2^2}}
\\
\qquad{}
=\lim_{s\rightarrow\infty}\frac{2z_1^4+\big({-}2-s^{-1}+(2s^2)^{-1}\big)z_1^4}{2s^{-1}z_1^2}=-\frac{z_1^2}{2}.
\end{gather*}
On the other hand
\begin{gather*}
\lim_{z_2\rightarrow0}\lim_{g\rightarrow2}\lim_{s\rightarrow\infty}\frac{\partial(\Delta E)}{\partial g}
=\lim_{z_2\rightarrow0}\lim_{g\rightarrow2}\lim_{s\rightarrow\infty}
\frac{g\big(z_1^2+z_2^2\big)^2-\big(2-(1-g+g(4s)^{-1})s^{-1}\big)\big(z_1^2-z_2^2\big)^2}
{4\sqrt{(1+g(4s)^{-1}-g/2)^2(z_1^2-z_2^2)^2+g^2z_1^2z_2^2}}
\\
\hphantom{\lim_{z_2\rightarrow0}\lim_{g\rightarrow2}\lim_{s\rightarrow\infty}\frac{\partial(\Delta E)}{\partial g}}{}
=\lim_{z_2\rightarrow0}\lim_{g\rightarrow2}\frac{g\big(z_1^2+z_2^2\big)^2-2\big(z_1^2-z_2^2\big)^2}
{4\sqrt{(1-g/2)^2(z_1^2-z_2^2)^2+g^2z_1^2z_2^2}}
\\
\hphantom{\lim_{z_2\rightarrow0}\lim_{g\rightarrow2}\lim_{s\rightarrow\infty}\frac{\partial(\Delta E)}{\partial g}}{}
=\lim_{z_2\rightarrow0}\frac{2\big(z_1^2+z_2^2\big)^2-2\big(z_1^2-z_2^2\big)^2}{4\sqrt{4z_1^2z_2^2}}
=\lim_{z_2\rightarrow0}{{z_1z_2}}=0.
\end{gather*}
This shows that the non-commutativity of limits is not just an artefact of the mean-f\/ield approximation,
but is an intrinsic property of the model.
Similar calculations also show
\begin{alignat*}{3}
& \lim_{s\rightarrow\infty}\lim_{g\rightarrow2}\lim_{z_2\rightarrow0}\frac{\partial(\Delta E)}{\partial g}
=-\frac{z_1^2}{2},
\qquad &&
\lim_{z_2\rightarrow0}\lim_{s\rightarrow\infty}\lim_{g\rightarrow2}
\frac{\partial(\Delta E)}{\partial g}=0,&
\\
& \lim_{g\rightarrow2^+}\lim_{z_2\rightarrow0}\lim_{s\rightarrow\infty}\frac{\partial(\Delta E)}{\partial g}
=\frac{z_1^2}{2},
\qquad &&
\lim_{g\rightarrow2^-}\lim_{z_2\rightarrow0}\lim_{s\rightarrow\infty}
\frac{\partial(\Delta E)}{\partial g}=-\frac{z_1^2}{2},&
\\
& \lim_{g\rightarrow2^+}\lim_{s\rightarrow\infty}\lim_{z_2\rightarrow0}\frac{\partial(\Delta E)}{\partial g}
=\frac{z_1^2}{2},
\qquad &&
\lim_{g\rightarrow2^-}\lim_{s\rightarrow\infty}\lim_{z_2\rightarrow0}
\frac{\partial(\Delta E)}{\partial g}=-\frac{z_1^2}{2}. &
\end{alignat*}

\section{Indicators of quantum phase transitions}
\label{numeric}

In the remaining subsections we compute various quantities which accord with the phase diagram shown in
Fig.~\ref{fig2}.
While there have been numerous studies of quantum phase transitions, there still lacks a~def\/initive
property which may be used to identify them.
Using numerical diagonalisation of the Hamiltonian~\eqref{H}, below we compute the energy gap, f\/idelity,
and entanglement (as given by the von Neumann entropy).
Each quantity is plotted as a~function of the parameter~$g^{-1}$.
We f\/ix $\mu=-0.99$, $z_1=1$, and adopt the parameterisation $z_2=(2000s)^{-1/2}$ such that
$z_2\rightarrow 0$ as $s\rightarrow\infty$.
We focus on this case as it permits us to compactly cover the regions~II,~III and~IV in Fig.~\ref{fig2}.

\subsection{Energy gap}

\begin{figure}[t] \centering
\includegraphics[height=6.2cm]{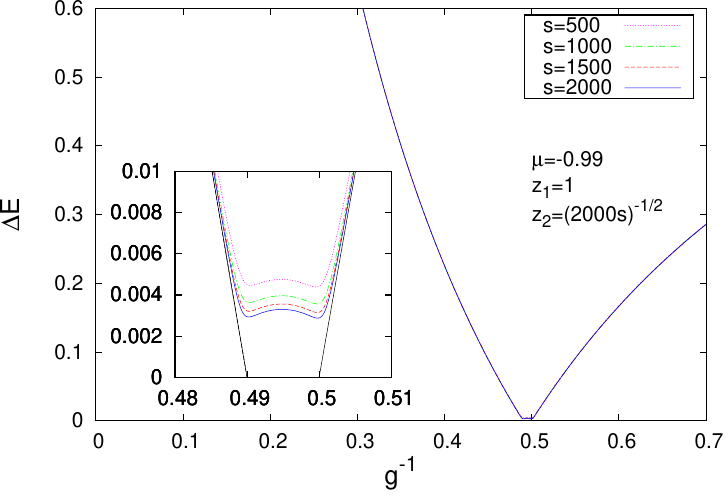}

\caption{Energy gap between the ground
state and the f\/irst excited state as a~function of~$g^{-1}$, with $\mu=-0.99$.
The results shown in the inset indicate the tendency towards vanishing gap in region~IV as $s\rightarrow
\infty$.
The solid black line represents the analytic curve for the energy gap as given by
equations~\eqref{gap2},~\eqref{gap3} for when $z_2=0$.}
\label{Gap}
\end{figure}

In Fig.~\ref{Gap} we plot the energy gap.
There are two dif\/ferent points of minima corresponding to the boundary lines between region III and IV,
and between region IV and II, in Fig.~\ref{phasez}.
It is seen in the inset that as $s$ increases the energy gap decreases in region IV of Fig.~\ref{fig2}.
The results appear more consistent with Fig.~\ref{fig2} rather than Fig.~\ref{fig1}, as to be expected in
view of the results from Subsection~\ref{muminusone}.

\subsection{Fidelity}

Another means to investigate quantum phase transitions is f\/idelity~\cite{zp06,zb07}, a~concept widely
used in quantum information theory.
The f\/idelity of two states is simply def\/ined as the modulus of the inner product
\begin{gather*}
\mathcal{F}(\psi,\phi)=|\langle\psi|\phi\rangle|.
\end{gather*}
Computing this quantity for ground-state vectors associated with Hamiltonians dif\/fering only by a~small
change in the coupling parameter, the f\/idelity may be sensitive to variations of the coupling parameter.
Generally a~minimum in the f\/idelity points to a~rapid change in the ground-state properties, which may
indicate a~critical point of the system.

\begin{figure}[t] \centering
 \includegraphics[height=6.2cm]{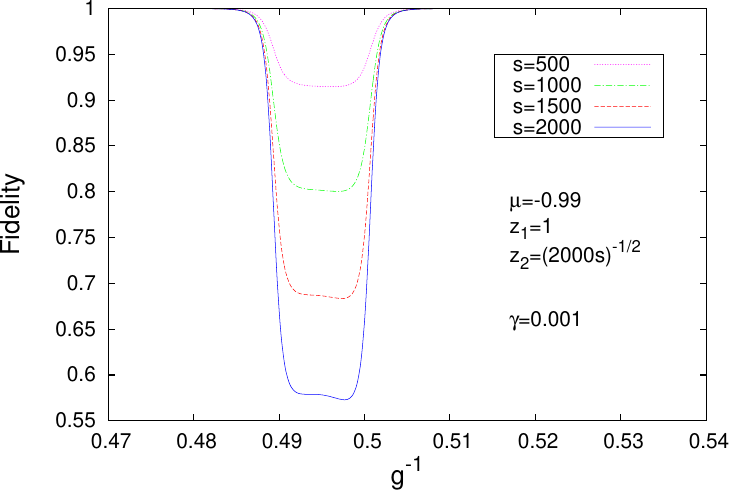}

 \caption{Ground-state f\/idelity as
a~function of~$g^{-1}$, with $\mu=-0.99$ and $\gamma=0.001$ for dif\/ferent values of $s$ as indicated in
the legend.
The results show signif\/icantly decreasing f\/idelity in region~IV as $s\rightarrow \infty$.}\label{Fid}
\end{figure}

Here we take two coupling values
\begin{gather*}
\big(g^{\pm}\big)^{-1}=g^{-1}(1\pm\gamma)
\end{gather*}
with $\gamma=0.001$ to compute the f\/idelity of the ground states associated with $g^{\pm}$, and then
consider the behaviour of the f\/idelity as $g$ is varied.
As is seen in Fig.~\ref{Fid}, for values of the parameter~$g^{-1}$ above~$0.5$ and values below~$0.49$, the
f\/idelity is close to~1, but suf\/fers an abrupt change at these two points.
The two points correspond to the boundary lines between regions~II and~IV and regions~IV and~III for
$\mu=-0.99$ in the phase diagram Fig.~\ref{phasez}.
Also, it can be seen that the minimum of the f\/idelity is decreasing as~$s$ increases.

\subsection{Entanglement}

\begin{figure}[t] \centering
\includegraphics[height=6.2cm]{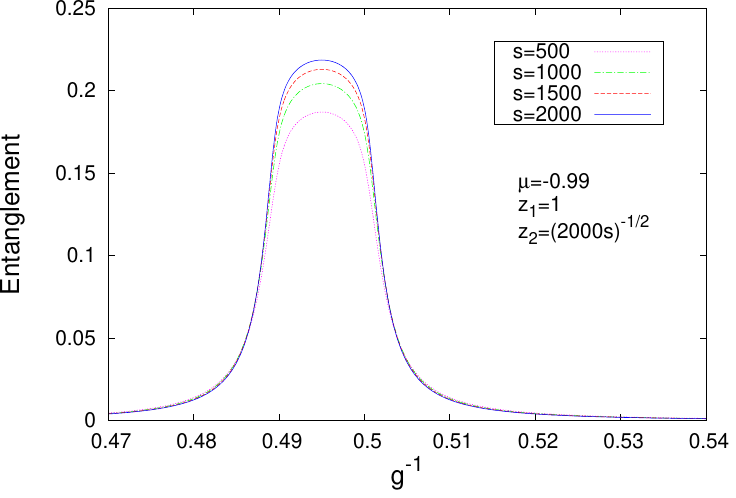}

\caption{Ground-state entanglement as
function of $g^{-1}$ for $\mu=-0.99$ for dif\/ferent values of $s$ as indicated in the legend.
The results show signif\/icantly increasing entanglement in region~IV as $s\rightarrow \infty$.}
\label{Ent}
\end{figure}

The coupled-spin model is naturally seen as a~bipartite system.
In this case we can use the von Neumann entropy of the reduced density operator to quantify the
entanglement of the system~\cite{hmm03,on02,pd05,rfmr12}.
Using the fact that the total spin $m$ is conserved, we can write a~general state of the system (for $m\leq
0$) in terms of~\eqref{basis} by
\begin{gather}
|\Psi\rangle=\sum_{j=-s}^{m+s}d_j|s,j\rangle\otimes|s,m-j\rangle,
\label{gs}
\end{gather}
for some complex numbers $d_j$.
The density operator for the state~\eqref{gs} is given by
\begin{gather*}
\rho=|\Psi\rangle\langle\Psi|=\sum_{i,j=-s}^{m+s}
d_id^*_j|s,i\rangle\langle s,j|\otimes|s,m-i\rangle\langle s,m-j|.
\end{gather*}
We take the partial trace with respect to the spin sector~$2$, yielding the reduced density operator for
the spin sector $1$:
\begin{gather*}
\rho_1=\mathrm{tr}_2(\rho)=\sum_{j=-s}^{m+s}|d_j|^2|s,j\rangle\langle s,j|.
\end{gather*}
The von Neumann entropy of entanglement for the state is given by
\begin{gather}
{\mathcal E}(\rho_1)=-\mathrm{tr}[\rho_1\log_2(\rho_1)]=-\sum_{j=-s}^{m+s}|d_j|^2\log_2\big(|d_j|^2\big).
\label{entg}
\end{gather}
Using the formula~\eqref{entg} we compute the entanglement of the ground state with $\mu=-0.99$ as function
of $g^{-1}$, as shown in Fig.~\ref{Ent}.
We can identify two abrupt changes in the entanglement value about the points $g^{-1}=0.49$ and
$g^{-1}=0.5$ in agreement with energy gap and f\/idelity analysis.
Note that the entanglement variation gets sharper as the value of~$s$ is increased.

\section{Conclusion}
\label{conc}

A main outcome of this study is that the limiting behaviours of the Hamiltonian~\eqref{H} depend on the
order in which limits are taken.
This was f\/irst observed through the mean-f\/ield analysis conducted in Section~\ref{mfa}.
In the following Section~\ref{exactsol}, concerned with exact techniques, two phase diagrams were
obtained as depicted in Figs.~\ref{fig1} and~\ref{fig2}.
Subsequent exact calculations, undertaken in Subsection~\ref{muminusone}, gave weight to the phase diagram
of Fig.~\ref{fig2}.
This was further supported by numerical calculation of the energy gap, f\/idelity, and entanglement,
conducted in Section~\ref{numeric}.
The discrepancy between the two diagrams has its origin in assumptions regarding limits commuting,
particularly those involving $z_2\rightarrow 0$.
In part, this can be traced back to the vanishing of the commutator~\eqref{magic} as $z_2\rightarrow 0$.
The calculations of the partial derivative of the energy gap with respect to the coupling~$g$ given in
Subsection~\ref{muminusone} provided concrete examples of non-commutativity of limits.

The model studied here may be viewed as an analogue of the $p+ip$ pairing Hamiltonian.
Letting $c_{{\bf k}}$, $c_{{\bf k}}^{\dagger}$ denote annihilation and creation operators for
two-dimensional spinless or spin-polarised fermions of mass $m$ with momentum ${\bf k}=(k_{x},k_{y})$, the
Hamiltonian reads~\cite{dilsz10,ilsz09,rdo10}
\begin{gather*}
H=\sum_{{\bf k}}\frac{|\bf{k}|^{2}}{2m}c^\dagger_{{\bf k}}c_{\bf{k}}-\frac{G}{4m}\sum_{{\bf k}\neq{\bf k}'}
(k_{x}+ik_{y})(k'_{x}-ik'_{y})c_{{\bf k}}^{\dagger}c_{-{\bf k}}^{\dagger}c_{-{\bf k}'}c_{{\bf k}'},
\end{gather*}
where $G$ is the coupling constant.
The Hamiltonian commutes with the total fermion number
\begin{gather*}
N=\sum_{{\bf k}}c^\dagger_{{\bf k}}c_{\bf{k}}.
\end{gather*}
This model has attracted interest due to it topological properties~\cite{bdh09,dr11,lcs11,rg00,sc06,v}.
The phase diagram of the model is characterized by two lines known as the Moore--Read line and the
Read--Green line.
Letting $L$ denote the number of momenta ${\bf k}$, the Moore--Read line is given by $G^{-1}=L-N/2$, for
which the ground state energy is zero.
The Read--Green line is given by $G^{-1}=(L-N/2)/2$, which was shown in~\cite{rdo10} to correspond to
a~third-order quantum phase transition.
Equation~\eqref{mrl} is the analogue of the Moore--Read line for the coupled spin Hamiltonian~\eqref{H},
while equation~\eqref{rgl} is the analogue of the Read--Green line.
The phase diagram for the $p+ip$ pairing Hamiltonian has the same qualitative features as Fig.~\ref{fig1}.
The topological nature of this model is ref\/lected by the fact that the equations for the Moore--Read and
Read--Green lines are independent of the distribution of the momenta ${\bf k}$, so the phase diagram is
robust against changes to the momentum distribution.
Analogously, equations~\eqref{mrl} and~\eqref{rgl} do not depend on the variables $z_1$ and $z_2$ for the
coupled-spin model, which may be viewed as a~degenerate limiting case of the $p+ip$ pairing Hamiltonian as
discussed in~\cite{ml12}.
Our studies indicate that new features of the phase diagram appear in the limit $z_2\rightarrow 0$.
This observation prompts a~closer examination of the $p+ip$ model in the limit when some of the momenta go
to zero, particularly with respect to the commutivity or otherwise of particular limits.
This aspect is signif\/icant because the occurrence of zero momenta in the $p+ip$ model is intimately
related to the existence of Majorana fermions, exhibiting non-abelian anyonic statistics, a~topic which has
been studied in~\cite{cld12,gr07,kafs09,rg00,tdnzz07}.

\subsection*{Acknowledgements}

EM is supported by CAPES (Brazil).
JL is supported by the Australian Research Council through Discovery Project DP110101414.
We thank the anonymous referees for their recommendations on revising the manuscript.
JL thanks Germ\'an Sierra for numerous insightful discussions.

\pdfbookmark[1]{References}{ref}
\LastPageEnding

\end{document}